\documentclass[aps,preprint,showpacs]{revtex4}
\usepackage{graphicx}

\begin{document}
\newcommand{\be}{\begin{equation}}
\newcommand{\ee}{\end{equation}}
\newcommand{\bea}{\begin{eqnarray}}
\newcommand{\eea}{\end{eqnarray}}

\title{Density of Yang-Lee zeros for the Ising ferromagnet}
\author{Seung-Yeon Kim}
\email{sykim@ssu.ac.kr}
\affiliation{Computer-Aided Molecular Design Research Center,
Soongsil University, Seoul 156-743, Korea}

\date{\today}

\begin{abstract}
The densities of Yang-Lee zeros for the Ising ferromagnet on the $L\times L$ square lattice
are evaluated from the exact grand partition functions ($L=3\sim16$).
The properties of the density of Yang-Lee zeros are discussed
as a function of temperature $T$ and system size $L$.
The three different classes of phase transitions for the Ising ferromagnet,
first-order phase transition, second-order phase transition,
and Yang-Lee edge singularity, are clearly distinguished by estimating
the magnetic scaling exponent $y_h$ from the densities of zeros for finite-size systems.
The divergence of the density of zeros at Yang-Lee edge in high temperatures
(Yang-Lee edge singularity), which has been detected only by the series expansion until now
for the square-lattice Ising ferromagnet, is obtained from the finite-size data.
The identification of the orders of phase transitions in small systems is
also discussed using the density of Yang-Lee zeros.
\end{abstract}


\maketitle


\section{Introduction}

Yang and Lee \cite{yang} proposed a mechanism
for the occurrence of phase transitions in the thermodynamic limit
and yielded an insight into the problem of the
Ising ferromagnet at arbitrary temperature in an arbitrary nonzero external magnetic field
by introducing the concept of the zeros of the grand partition function $Z(T,H)$
(for fluid systems, $Z(T,\mu)$)
in the {\it complex} magnetic-field (for fluid systems, fugacity) plane (Yang-Lee zeros).
They \cite{lee} also formulated
the celebrated circle theorem which states that
Yang-Lee zeros of the Ising ferromagnet lie on the unit circle $x_0=e^{i\theta}$
in the complex $x=e^{-2H/k_B T}$ plane.
Below and at the critical temperature $T_c$, Yang-Lee zeros cut the positive real axis
at the ferromagnetic transition point $x_c=1$ ($H_c=0$) in the thermodynamic limit.
The spontaneous magnetization $m_0$ is determined by the density of Yang-Lee zeros
$g(\theta)$ on the positive real axis, i.e., $m_0=2\pi g(\theta=0)$.
However, above the critical temperature, Yang-Lee zeros do not cut the positive real axis
in the thermodynamic limit. There is a gap in the distribution of zeros
around the positive real axis, that is, $g(\theta)=0$ for $|\theta|<\theta_e$.
Within this gap, the free energy is analytic and there is no phase transition.
The Yang-Lee zeros at $\theta=\pm\theta_e$ are called the Yang-Lee edge zeros whose
locations have never been known exactly for $T>T_c$ and $d\ge2$.
At the critical temperature the gap disappears, i.e., $\theta_e=0$.
As temperature changes from $T_c$ to $\infty$, the Yang-Lee edge $\theta_e$
moves from $\theta=0$ to $\theta=\pi$ \cite{kortman,creswick97,wang,kim98}.

In addition to Yang-Lee zeros,
Fisher showed that the thermal properties of the square-lattice Ising model for $H=0$
are completely determined by the zeros of the canonical partition function $Z(T)$
in the complex temperature plane (Fisher zeros) \cite{fisher65,lu}.
Until now, Yang-Lee zeros and Fisher zeros
(and other kinds of zeros for some problems)
of numerous physical, chemical, biological, and mathematical systems
have been investigated extensively to understand their important properties \cite{bena}.

If the density of zeros
\cite{lee,kortman,creswick97,fisher65,lu,bena,suzuki70,kenna94,kenna97,janke02,binek,binek2,kim04a,suzuki67}
is found, the free energy, the equation of state,
and all other thermodynamic functions can be obtained.
However, very little is known about the actual form of the density of zeros.
The density of Yang-Lee zeros for the Ising ferromagnet has never been known exactly
except for the one-dimensional case \cite{lee}.
Suzuki {\it et al.} \cite{suzuki70} obtained the exact grand
partition functions of the Ising model on the $4\times6$ square lattice and
the $3\times3\times3$ simple-cubic lattice, and calculated
the densities of Yang-Lee zeros for these small-size lattices.
Kortman and Griffiths \cite{kortman} investigated
the densities of Yang-Lee zeros for the Ising ferromagnet
on the square lattice and the diamond lattice,
based on the high-field, high-temperature series expansion,
and found that the density of zeros for the square-lattice Ising ferromagnet
diverges at Yang-Lee edge $\theta_e$ in high temperatures.
For the $L\times L$ square-lattice Ising model,
Creswick and Kim \cite{creswick97} obtained the exact grand partition functions up to $L=10$
and the semi-exact grand partition functions up to $L=14$,
calculated the densities of Yang-Lee zeros for $T\le T_c$,
and evaluated the spontaneous magnetization $m_0(T)=2\pi g(\theta=0,T)$
in the infinite-size limit.
From Monte Carlo data, the density of Yang-Lee zeros
near the transition point has been studied
for the four-dimensional Ising ferromagnet \cite{kenna94} and
the square-lattice XY ferromagnet \cite{kenna97,janke02}.
The density of Yang-Lee zeros has also been investigated experimentally
for the two-dimensional Ising ferromagnet FeCl$_2$ in axial magnetic fields \cite{binek,binek2}.

Until now, the divergence of the density of zeros at Yang-Lee edge
(the so-called Yang-Lee edge singularity)
for the square-lattice Ising ferromagnet in high temperatures has been obtained only by
the series expansion \cite{kortman,kurtze1,baker86}.
Furthermore, for $T>T_c$, the finite-size effects of the density of Yang-Lee zeros
for the Ising ferromagnet have never been studied.
In this paper, we investigate the properties of the density of Yang-Lee zeros
based on the exact grand partition functions of the $L\times L$ square-lattice Ising model
($L=3\sim16$).
We deal with the three different classes of phase transitions for the Ising ferromagnet,
(1) first-order phase transition, (2) second-order phase transition,
and (3) Yang-Lee edge singularity, in a unified framework using the densities of zeros
for finite-size systems.
We also study the density of Yang-Lee zeros
and its divergence in high temperatures from the finite-size data.


\section{Density of zeros}

For a finite-size system of size $L$, the density of Yang-Lee zeros $g(\theta,L)$
at the critical temperature $T_c$ is given by \cite{creswick97,bena}
\be
g(\theta,L)=L^{-d+y_h}g(\theta L^{y_h})
\ee
near $\theta=0$.
Equation (1) implies
\be
g(\theta)\sim|\theta|^{(d-y_h)/y_h}=|\theta|^{1/\delta}
\ee
for the infinite system.
At the critical temperature the value of the magnetic scaling exponent $y_h$
for the Ising ferromagnet in two dimensions ($d=2$)
is well known to be 15/8, that is, $\delta=15$.
Therefore, the density of zeros vanishes at the critical point $H_c=0$ ($\theta=0$).
This kind of behavior of the density of zeros at $T_c$ is the characteristic of
a second-order phase transition.
On the other hand, for $T<T_c$, we have the finite density of zeros at $\theta=0$,
$g(0)=m_0/2\pi$, which is the characteristic of a first-order phase transition.

To describe the density of zeros {\it above} the critical temperature as well,
we propose that Eqs.~(1) and (2) can be generalized as
\be
g(\Theta,L)=L^{-d+y_h}g(\Theta L^{y_h})
\ee
and
\be
g(\Theta)\sim|\Theta|^{(d-y_h)/y_h},
\ee
where $\Theta=\theta-\theta_e$.
For $T\le T_c$, Eq.~(3) is the same as Eq.~(1) because of $\theta_e=0$.
The form $g(\Theta)\sim|\Theta|^\sigma$ was suggested by Suzuki \cite{suzuki67}
who assumed $\sigma=1/\delta={1\over15}$ for the two-dimensional Ising ferromagnet
even above the critical temperature.
According to Suzuki's assumption, the density of zeros above the critical temperature
vanishes at Yang-Lee edge $\theta_e$.

Using the high-temperature series expansion for the Ising ferromagnet on the square lattice,
Kortman and Griffiths \cite{kortman} showed that the density of Yang-Lee zeros,
$g(\Theta)\sim|\Theta|^\sigma$,
diverges with $\sigma=-0.12(5)$ at $\theta=\theta_e$ for $T > 2T_c$,
in disagreement with Suzuki's assumption.
Kurtze and Fisher \cite{kurtze1} refined the estimation of $\sigma$ by analyzing
the high-temperature series expansion for the classical $n$-vector model
and the quantum Heisenberg model in the limit of infinite temperature,
and reported $\sigma=-0.163(3)$ in two dimensions (square and triangular lattices).
Baker {\it et al.} \cite{baker86} analyzed the series expansion of much greater length for
the square-lattice Ising ferromagnet, and obtained $\sigma=-0.1560(33)$
at $y=e^{-2\beta J}={2\over3}$ ($T\approx 5 J/k_B$) and $\sigma=-0.1576(34)$
at $y={4\over5}$ ($T\approx 9 J/k_B$) from integral approximants.
Remarkably, the exponent $\sigma$ has also been experimentally measured to be
$-0.15(2)$ in the range $49K\le T\le53K$ and $-0.365$ at $T=34K$
for the triangular-lattice Ising ferromagnet FeCl$_2$ \cite{binek2}.

Fisher \cite{fisher78} renamed the edge zero as the Yang-Lee edge {\it singularity}
for $T>T_c$, and proposed the idea that the Yang-Lee edge singularity
can be thought of as a new second-order phase transition with associated critical exponents.
Above the critical temperature the value of $\sigma$ is known to be
$\sigma=-{1\over6}$ in two dimensions \cite{dhar},
resulting in $y_h={12\over5}$ (according to Eq.~(4)), clearly different from
the value of $y_h={15\over8}$ for the Ising ferromagnet at the critical temperature.
The study of the Yang-Lee edge singularity has been extended to
the spherical model \cite{kurtze2},
the quantum one-dimensional transverse Ising model \cite{uzelac},
the hierarchical model \cite{baker79},
branched polymers \cite{parisi},
the Ising models on fractal lattices \cite{southern},
the Ising systems with correlated disorder \cite{tadic},
fluid models with repulsive-core interactions \cite{lai},
and the antiferromagnetic Ising model \cite{kim05}, etc.


\section{grand partition function and number of states}

The Ising model in an external magnetic field $H$ on a lattice
with $N_s$ sites and $N_b$ bonds is defined by the Hamiltonian
\be
{\cal H}=J\sum_{\langle i,j\rangle}(1-\sigma_i\sigma_j)+H\sum_i(1-\sigma_i),
\ee
where $J$ is the coupling constant,
$\langle i,j\rangle$ indicates a sum over all nearest-neighbor pairs
of lattice sites, and $\sigma_i=\pm1$.
The grand partition function of the Ising model is
$$
Z=\sum_{\{ \sigma_n \}} e^{-\beta{\cal H}},
$$
where $\{ \sigma_n \}$ denotes a sum over $2^{N_s}$ possible spin configurations
and $\beta=(k_BT)^{-1}$.
If we define the number of states, $\Omega(E,M)$,
with a given energy
\be
E={1\over2}\sum_{\langle i,j\rangle}(1-\sigma_i\sigma_j)
\ee
and a given magnetization
\be
M={1\over2}\sum_i(1-\sigma_i),
\ee
where $E$ and $M$ are positive integers $0\le E\le N_b$ and $0\le M\le N_s$,
then the grand partition function can be written as
\be
Z(x,y)=\sum_{E=0}^{N_b}\sum_{M=0}^{N_s}\Omega(E,M) y^E x^M,
\ee
where $y=e^{-2\beta J}$ and $x=e^{-2\beta H}$.
From Eq.~(8) it is clear that $Z(x,y)$ is simply a polynomial in $x$ and $y$.

The {\it exact} integer values for the number of states $\Omega(E,M)$ of the Ising model
on the $L\times L$ square lattices with cylindrical boundary conditions have been evaluated
up to $L=10$ to study the density of Yang-Lee zeros for $T\le T_c$
using the microcanonical transfer matrix ($\mu$TM) \cite{creswick97}.
In Ref.~[4], for $11\le L\le14$, we had to evaluate
the coefficients, for a fixed value of $y$,
$$
\omega(M)=\sum_E\Omega(E,M)y^E
$$
as real numbers of finite precision due to memory limitations.
This semi-exact method worked well for $T\le T_c$, and produced all zeros on the unit circle.
However, for $T>T_c$ the number of zeros on the unit circle,
obtained from $Z(x)=\sum\omega(M)x^M$, decreases as $T$ increases.
Therefore, to study the density of Yang-Lee zeros for $T>T_c$ we need
the exact integer values for $\Omega(E,M)$, not $\omega(M)$.
In this work, the exact integer values for $\Omega(E,M)$ are calculated up to $L=16$
($2^{256}=1.158\times10^{77}$ configurations).
The memory requirement during $\mu$TM calculation
is $2\times4\times(N_s/2+1)\times(N_b+1)\times P_m\times 2^L$ byte
for the two-dimensional Ising model,
where $P_m$ is the maximum size for storing very long integer numbers.
For $L=16$, the memory requirement is $3.03\times10^{11}$ byte.
(For comparison, the memory requirement for $L=10$, the largest size calculated in Ref.~[4],
is $3.19\times10^8$ byte.
The largest size calculated before this work is $L=14$ whose memory requirement
is $3.44\times10^{10}$ byte \cite{kim04b}.)
The largest number of states for $L=16$ is
\bea
\Omega(249,128)=
207964931539339552089271840638951892633 \cr
537987437849943362668628644700590272,
\eea
which is approximately $2.08\times10^{74}$.


\section{Numerical results}

For a finite-system of size $L$, the density of zeros (per site) may be defined as
\be
g(\theta_k,L)={1\over L^d}{1\over{\theta_{k+1}-\theta_k}},
\ee
where $\theta_k$ ($k=1,2,3,...$) is the argument of the $k$-th zero on the
unit circle $x_0=e^{i\theta}$ of $Z(x,y)$.
The index $k$ is counted from $x=1$ ($\theta=0$) to $x=-1$ ($\theta=\pi$)
for the zeros with $\theta\ge0$. Hence, $\theta_1(L)=\theta_e(L)$.
Figure 1 shows the density of Yang-Lee zeros for the $16\times16$ Ising ferromagnet
at various temperatures (in unit of $J/k_B$).
The Yang-Lee zeros at $T=0$ are uniformly distributed on the unit circle \cite{kim98},
and the density of zeros is given by $g(\theta)=1/2\pi$ ($=0.159...$).
As shown in Fig.~1, for a low temperature $T=1$, the density of zeros is
nearly identical with $g(\theta)=0.159$.
The spontaneous magnetization can be evaluated from the finite-size data for the density
of zeros ($m_0(L)=2\pi g(0,L)$, $L=3\sim16$) using the BST extrapolation algorithm \cite{bst}.
For $T=1$ the extrapolated value of the spontaneous magnetization is
$m_0=0.9992757(3)$ in excellent agreement with the exact result $0.9992757519...$ \cite{yang2}.
At the critical temperature $T_c$ ($=2.269...$), the maximum value of the density of zeros
is 0.166096 at $\theta=2.541$ where the density of zeros begins to decrease monotonically
as $\theta$ decreases.
At $T_c$ the density of zeros vanishes at $\theta=0$ in the limit $L\to\infty$,
according to Eq.~(2).
For $T=3$ the density of zeros is similar to that at $T_c$, except for
a violation of the monotonic decrease of the density of zeros around $\theta=0.2$.
It is interesting that the minimum value of the density of zeros at $T=3$
($g(\theta_e=0.126)=0.112$) is lower than that at $T_c$ ($g(0.023)=0.117$).
Therefore, we may guess that the density of zeros for $T$ just above $T_c$, for example $T=3$,
vanishes at Yang-Lee edge $\theta_e$ in the infinite-size limit.
As shown in Fig.~1, at high temperatures such as $T=5$ and 10, the overall density of zeros
becomes higher because the Yang-Lee edge $\theta_e$ tends to $\pi$ as $T$ rises.
For $T=5$ the density of zeros has a local maximum $g=0.182$ near $\theta_e=0.532$.
Finally, for $T=10$, the density of zeros has the global maximum $g=0.279$
near $\theta_e=1.161$.
This fact may be a precursor for the divergence of the density of zeros at $\theta_e$
in the infinite-size limit for high temperatures.

Figure 2 shows the densities of Yang-Lee zeros for $L=12$ and 16 at different temperatures.
Well below the critical temperature (for example, $T=1$), the densities of zeros
for $L=12$ and 16 are not distinguishable, as shown in Fig.~2(a).
Therefore, the finite-size effect for the density of zeros is very small
well below $T_c$.
At and near the critical temperature (for example, $T=2.5$ in Fig.~2(a)),
the densities of zeros for $L=12$ and 16 are also indistinguishable,
except for them near $\theta=0$ where the density of zeros $g(L)$ decreases sharply
as the system size $L$ increases.
Well above the critical temperature (for example, $T=10$ in Fig.~2(b)),
the overall forms of the density of zeros for $L=12$ and 16 are almost identical.
The finite-size effect for the density of zeros at $T=10$ is relatively large,
compared to those at $T=1$ and 2.5.
Near the Yang-Lee edge $\theta_e=1.161$, the density of zeros is somewhat irregular,
and it increases as $L$ increases.

If we set $\Theta=0$ ($\theta=\theta_e(L)$) in Eq.~(3), we have
\be
g(0,L)=L^{-d+y_h}g(0),
\ee
from which we can define the magnetic scaling exponent
\be
y_h(L)=d+{{{\rm ln}[g(0,L+1)/g(0,L)]}\over{\rm ln}[(L+1)/L]}
\ee
for finite lattices.
Table I shows the values of $y_h(L)$ at $T=1$, $T_c$, and 10, respectively.
For $T=1$, $y_h(L)$ changes very slightly around $d=2$ as $L$ increases.
This kind of behavior may result from the small finite-size effect
for the density of zeros well below the critical temperature.
At $T_c$ and $T=10$, $y_h(L)$ increases monotonically as $L$ increases.
In particular, at $T=10$, $y_h(L)$ increases quickly.

Table II shows the extrapolated results for the magnetic scaling exponent $y_h$
by the BST algorithm.
For $T=1$, the extrapolated result clearly indicates $y_h=d=2$ that is expected
at a strong first-order phase transition \cite{fisher82}.
At $T_c$, the extrapolated value of $y_h=1.8747(7)$ is in agreement with
the exact value of $y_h={15\over8}=1.875$.
For $T\ge3$, we obtained the extrapolated value of $y_h=2.4$
which gives $\sigma=-{1\over6.0}$ (the characteristic of Yang-Lee edge singularity).
According to Eq.~(4), the density of zeros diverges at Yang-Lee edge $\theta_e$
when $y_h>d$.
Finally, we can easily distinguish the three different classes of phase transitions
(first-order phase transition, second-order phase transition,
and Yang-Lee edge singularity)
by estimating $y_h$ from the densities of zeros on finite lattices.

Near the critical temperature $T_c=2.269...$, we expect some crossovers
between different classes.
At $T=2.2$ we obtained $y_h=1.98(15)$ that is close to $y_h=2$
but is not clearly distinguished from $y_h={15\over8}$ due to the large error.
At $T=2.4$ and 2.5, the extrapolated values for $y_h$ are 2.33(23) and 2.89(72),
respectively, indicating the existence of Yang-Lee edge singularity.
But their errors are too large.
It is interesting that the densities of zeros for finite-size systems at $T=2.4$, 2.5, and 3
are quite similar to those at $T_c$
but they give the values of $y_h$ ($>d=2$) completely different
from $y_h={15\over8}$ ($<d$).
For the first time, it is shown that, even in the range of $2.4\le T \le 4$,
the square-lattice Ising ferromagnet has the Yang-Lee edge singularity.
At $T=2.3$ we obtained $y_h=1.58(4)$ whose nearest value is $y_h={15\over8}$ at $T_c$.
This value of $y_h=1.58(4)$ implies that the density of zeros for $T>T_c$
vanishes at $\theta_e$.
However, if the data of much larger sizes can be used,
there will be a possibility of obtaining $y_h > 2$ even at $T=2.3$.

The finite-size data for the density of zeros $g(\theta_e,L)$
($L=3\sim16$) in high temperatures have also been extrapolated to infinite size.
In high temperatures, it is convenient to deal with the inverse density of zeros
$f(\theta_e,L)\equiv1/g(\theta_e,L)$ because
$f(\theta_e,L)$ vanishes with the scale factor $L^{-(y_h-d)}$, in the limit $L\to\infty$,
but $g(\theta_e,L)$ diverges.
Table III shows the extrapolated results for the inverse density of zeros $f(\theta_e)$.
For $T\ge5$ the extrapolated results imply the divergence of the density of zeros
at $\theta_e$.
However, at $T=4$, we can not conclude anything only from the extrapolated result
for the inverse density of zeros because of the large error.


\section{phase transitions in small systems}

In the previous section, the density of zeros has been extracted
from the locations of Yang-Lee zeros for finite sizes,
using Eq.~(10) \cite{creswick97,suzuki70,kim04a}.
Recently, two alternative approaches (JK approach \cite{janke02,janke01,janke04,alves}
and BMH approach \cite{borrmann,alves}) have been proposed
to extract the density of zeros from the locations of Fisher zeros
in the complex temperature ($y=e^{-2\beta J}$) plane for finite sizes
and to study phase transitions in small systems.
These two approaches are based on the assumption that Fisher zeros
$y_j=s_j+i t_j=y_c+i r_j\exp(-i\psi)$
close to the critical point $y_c$ lie on a single line for quite large $L$,
where the index $j(=1,2,3,...)$ increases with distance $r_j$ from the critical point,
$s_j={\rm Re}(y_j)=y_c+r_j\sin\psi$, and $t_j={\rm Im}(y_j)=r_j\cos\psi$.

In the JK approach \cite{janke02,janke01,janke04,alves}, the density of zeros is defined as
\be
\lambda_L(r)=L^{-d}\sum_j\delta(r-r_j(L)),
\ee
and the cumulative density of zeros $\Lambda_L(r)\equiv\int_0^r \lambda_L(r')dr'$
becomes a step function with
\be
\Lambda_L(r)={j\over L^d}\ \ \ [{\rm if}\ r\in(r_j,r_{j+1})]
\ee
for a finite-size system of size $L$. Then the average cumulative density of zeros
is given by
\be
\Lambda_L(r_j)=\frac{2j-1}{2L^d},
\ee
and it is expressed as
\be
\Lambda_{\infty}(r) = \lambda_{\infty}(0)r + a r^{w+1} + \cdots
\ee
in the limit $L\to\infty$.
Here, the density of zeros $\lambda_{\infty}(0)$ on the real axis is proportional
to the latent heat $\Delta e$.
At a first-order phase transition, the first term $\lambda_{\infty}(0)r$ becomes
dominant for small $r$.
At a second-order phase transition, the second term $a r^{w+1}$
is the leading one because of $\lambda_{\infty}(0)=0$.
Finally, the cumulative density of zeros can be approximated as
\be
\frac{2j-1}{2L^d}=\Lambda_L(r_j)=a_1 [t_j(L)]^{a_2} + a_3
\ee
because the parameter $r_j$ for Fisher zeros close to the critical point $y_c$
may be taken to be the imaginary part $t_j(L)$ due to $y_c\sim s_j(L)$.
A value of $a_3$ inconsistent with zero indicates the absence of a phase transition.
The parameter $a_2$ determines the order of a phase transition.
A first-order phase transition takes a value $a_2 \sim 1$ for small $t$,
whereas a value of $a_2$ larger than $1$ indicates a second-order phase transition
whose specific-heat exponent is given by $\alpha=2-a_2$.
For a first-order phase transition, the parameter $a_1$ corresponds to
the density of zeros $\lambda_{\infty}(0)$ on the real axis.
The parameter $a_2$ is also expressed as $a_2=d/y_t$
in terms of the thermal scaling exponent $y_t$.
The JK approach has also been applied to the Yang-Lee zeros of the square-lattice
XY ferromganet \cite{janke02} where the parameter $a_2$ is expressed as $a_2=d/y_h$
in terms of the magnetic scaling exponent $y_h$,
that is,
\be
a_2={2d\over d+2-\eta}.
\ee
The JK approach has also been generalized to study
the areal distributions of Fisher zeros \cite{janke04}.

On the other hand, there are three important parameters
in the BMH approach \cite{borrmann,alves}.
The first parameter $t_1$ (the imaginary part of the first zero)
becomes 0 in the limit $L\to\infty$
for a true phase transition in the Ehrenfest sense \cite{creswick97,kim98,kim04b,kims1,kims2}.
If there is a phase transition, we assume that the line density of zeros,
\be
\phi(t_j)\equiv{1\over2}\left({1\over|y_j-y_{j-1}|}+{1\over|y_{j+1}-y_j|}\right),
\ee
follows a simple power law $\phi(t)\sim t^{\alpha_\phi}$ for small $t$.
The second parameter $\alpha_\phi$ can be estimated as
\be
\alpha_\phi={\ln\phi(t_3)-\ln\phi(t_2)\over\ln t_3-\ln t_2}
\ee
from the first four Fisher zeros ($y_1, y_2, y_3, y_4$).
The line of Fisher zeros also crosses the real axis
at the critical point $y_c$ with the angle $\psi$, yielding
the third parameter
\be
\gamma_\psi=\tan\psi={s_2-s_1\over t_2-t_1}.
\ee
The second and third parameters ($\alpha_\phi$ and $\gamma_\psi$) determine
the order of a phase transition.
The values of $\alpha_\phi=0$ ($\alpha_\phi$ can be less than zero for finite sizes)
and $\gamma_\psi=0$ indicate a first-order phase transition.
A second-order phase transition occurs for $0 < \alpha_\phi < 1$ and
arbitrary $\gamma_\psi$, and a higher-order phase transition for $\alpha_\phi > 1$
and arbitrary $\gamma_\psi$.
Alves {\it et al.} \cite{alves} have performed numerical comparisons
of the JK and BMH approaches for the Fisher zeros
of the two-dimensional four-state and five-state Potts models and of biological molecules,
using Monte Carlo data.
Their results implied that the JK approach is little better than the BMH approach.

The approach explained in the previous sections, which may be called as
the angular-density-of-zeros (ADOZ) approach, can also be used to study
phase transitions in small systems. In Eq.~(4), the density of Yang-Lee zeros
near $\Theta=0$ is given by $g(\Theta)\sim|\Theta|^{(d-y_h)/y_h}$.
Similarly, the density of Fisher zeros near $\Theta=0$
is also given by $g(\Theta)\sim|\Theta|^{(d-y_t)/y_t}$ \cite{kim04a}.
The exponent $(d-y_t)/y_t$ (or $(d-y_h)/y_h$) corresponds to the parameter
$a_2-1$ of the JK approach.
As in the BMH approach, the zero value of $t_1$
in the limit $L\to\infty$ indicates the existence of a true phase transition.
When there is no phase transition, Fisher (or Yang-Lee) edge singularity \cite{kims2,kim02}
exists if $y_t > d$ (or $y_h > d$).
As shown in Table I, the values of $y_h(L)$ at $T=10$ are clearly greater than $d=2$
for relatively large $L$, suggesting the existence of Yang-Lee edge singularity.
When there is a phase transition, the value of $y_t$ (or $y_h$) determines
the order of the phase transition.
To calculate $y_t(L)$ (or $y_h(L)$), we need only the first two Fisher (or Yang-Lee)
zeros according to Eqs.~(10) and (12).
A value of $y_t\approx d$ (or $y_h\approx d$) means a first-order phase transition.
The values of $y_h(L)$ at $T=1$ in Table I clearly indicate a first-order phase transition.
A second-order phase transition occurs for $d/2\le y_t<d$ (or $d/2\le y_h<d$),
and a higher-order phase transition for $y_t<d/2$ (or $y_h<d/2$).
The values of $y_h(L)$ at $T_c$ in Table I indicate a second-order phase transition.

Now the three (JK, BMH, and ADOZ) approaches are applied to Yang-Lee zeros
of the square-lattice Ising ferromagnet.
The BMH approach is applied to Yang-Lee zeros for the first time.
Because the JK and BMH approaches assume a line distribution of zeros
near a transition point, we use Yang-Lee zeros from $L=8$.
The JK and BMH approaches have been compared for the approximate zeros of
quite large (but still small) finite-size systems using Monte Carlo simulation \cite{alves}.
Here these approaches are compared for the exact zeros of very small finite-size systems
($L=8$ to 16).
Because the BMH approach requires the first four zeros, we use only them
in the applications of the JK and BMH approaches for a fair comparison.
The ADOZ approach requires the first two zeros of two different sizes,
in total, four zeros (see Eq.~(12)).

In the identification of the orders of phase transitions in small systems,
the most important quantities are $a_2(L)$, $\alpha_\phi(L)$, and $y_h(L)$,
respectively, for the JK, BMH, and ADOZ approaches.
Figure 3 shows the results of $a_2(L)$, $\alpha_\phi(L)$, and $y_h(L)$
at $T=1$ (first-order phase transition), $T=2.2$ (weak first-order phase transition),
and $T_c=2.269...$ (second-order phase transition) for the Yang-Lee zeros
of the square-lattice Ising ferromagnet.
Here, a weak first-order phase transition means a first-order phase transition
with very large (but finite) correlation length.
All the approaches clearly identify the first-order and second-order phase transitions.
In particular, the finite-size results at $T=1$ are perfect in that
they already reach the infinite-size values $a_2=1$, $\alpha_\phi=0$, and $y_h=2$.
But the three approaches fail to identify the weak first-order phase transition.
An advantage of the JK approach is that we can use the zeros from different sizes
together in the estimation of $a_2$. The only first Yang-Lee zeros from $L=13$ to 16
(in total, four zeros) are used to estimate $a_2$ at $T=2.2$ and $T_c$.
At $T_c$, we obtain $a_2=1.0874(4)$, close to the infinite-size value
$a_2=d/y_h=16/15(=1.0666...)$.
At $T=2.2$, the result is $a_2=1.011(2)$.
It is remarkable because the JK approach almost identifies
the weak first-order phase transition only from very small systems up to $L=16$.


\section{Conclusion}

We have enumerated the exact integer values for the number of states $\Omega(E,M)$ of the Ising model
on the $L\times L$ square lattice up to $L=16$, from which the grand partition function
$Z(T,H)=\sum_{E,M}\Omega(E,M) e^{-2\beta(J E+H M)}$ is obtained and
the density of Yang-Lee zeros $g(\theta,L)$ on the unit circle $x=e^{-2\beta H}=e^{i\theta}$
is evaluated for the Ising ferromagnet.
We have compared the densities of Yang-Lee zeros at different temperatures for $L=16$.
We have also discussed the finite-size effects for the densities of zeros at different temperatures.
The three different classes of phase transitions for the Ising ferromagnet,
(1) first-order phase transition, (2) second-order phase transition, and (3) Yang-Lee edge singularity
(the divergence of the density of zeros at Yang-Lee edge in high temperatures)
have been clearly distinguished by estimating
the magnetic scaling exponent $y_h$ from the densities of zeros on finite lattices.
The Yang-Lee edge singularity that has been detected only by the series expansion until now
for the square-lattice Ising ferromagnet, has also been obtained
using the densities of zeros for finite-size systems.
We have also discussed the identification of the orders of phase transitions
in small systems using the three (JK, BMH, and ADOZ) approaches to extract
the density of zeros from numerical data.


\begin{acknowledgments}
The author was supported by the Korea Research Foundation Grant
funded by the Korean Government (MOEHRD) (KRF-2005-005-J01103).
\end{acknowledgments}



\newpage

\begin{table}
\caption{The magnetic scaling exponent $y_h(L)$ for finite lattices,
obtained from the density of zeros $g(\theta_e,L)$ using Eq.~(12),
at $T=1$, $T_c$, and 10.}
\begin{ruledtabular}
\begin{tabular}{cccc}
$L$ &$T=1$   &$T_c$     &$T=10$    \\
\hline
3  &1.994240 &1.776068  &1.676531  \\
4  &1.999877 &1.795022  &1.819092  \\
5  &2.001264 &1.806392  &1.917883  \\
6  &2.001516 &1.814538  &1.992971  \\
7  &2.001475 &1.820774  &2.051768  \\
8  &2.001368 &1.825750  &2.098558  \\
9  &2.001256 &1.829834  &2.136280  \\
10 &2.001153 &1.833257  &2.167057  \\
11 &2.001062 &1.836174  &2.192455  \\
12 &2.000982 &1.838693  &2.213638  \\
13 &2.000913 &1.840891  &2.231483  \\
14 &2.000852 &1.842827  &2.246655  \\
15 &2.000798 &1.844547  &2.259666  \\
\end{tabular}
\end{ruledtabular}
\end{table}

\begin{table}
\caption{The extrapolated results for the magnetic scaling exponent $y_h$
at different temperatures.}
\begin{ruledtabular}
\begin{tabular}{cl}
$T$   &$y_h$ \\
\hline
1     &2.000000(6) \\
2.2   &1.98(15)    \\
$T_c$ &1.8747(7)   \\
2.3   &1.58(4)     \\
2.4   &2.33(23)    \\
2.5   &2.89(72)    \\
3     &2.39(3)     \\
4     &2.4056(11)  \\
5     &2.4005(3)   \\
10    &2.40026(5)  \\
15    &2.4001(2)   \\
20    &2.3998(3)   \\
\end{tabular}
\end{ruledtabular}
\end{table}

\begin{table}
\caption{The extrapolated results for the inverse density of zeros $f(\theta_e)$
at Yang-Lee edge $\theta_e$ in high temperatures.}
\begin{ruledtabular}
\begin{tabular}{rcc}
$T$ &$\theta_e$ &$f(\theta_e)$ \\
\hline
4   &0.314846(75) &0.27(756) \\
5   &0.514241(33) &$-0.01(7)$ \\
10  &1.149334(14) &$-0.05(4)$ \\
15  &1.479519(10) &$-0.02(4)$ \\
20  &1.687041(8)  &$-0.005(31)$ \\
\end{tabular}
\end{ruledtabular}
\end{table}


\begin{figure}
\includegraphics[width=15cm]{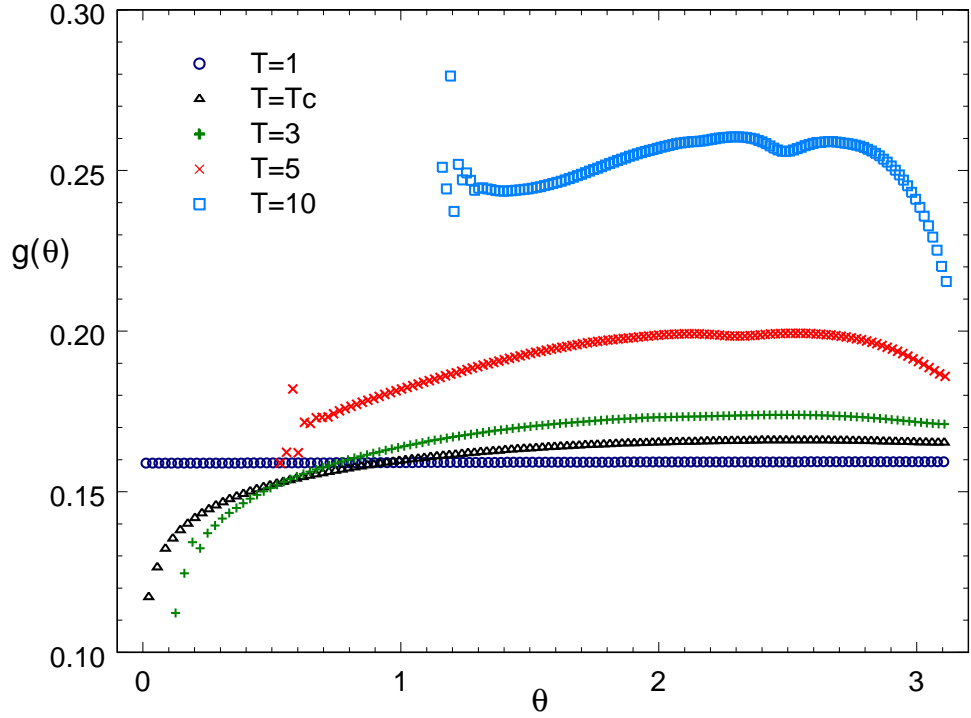}
\caption{(Color online) The density of Yang-Lee zeros for the $16\times16$ Ising ferromagnet
at $T=1$, $T_c$ ($=2.269...$), 3, 5, and 10 (in unit of $J/k_B$).}
\end{figure}

\begin{figure}
\includegraphics[width=14cm]{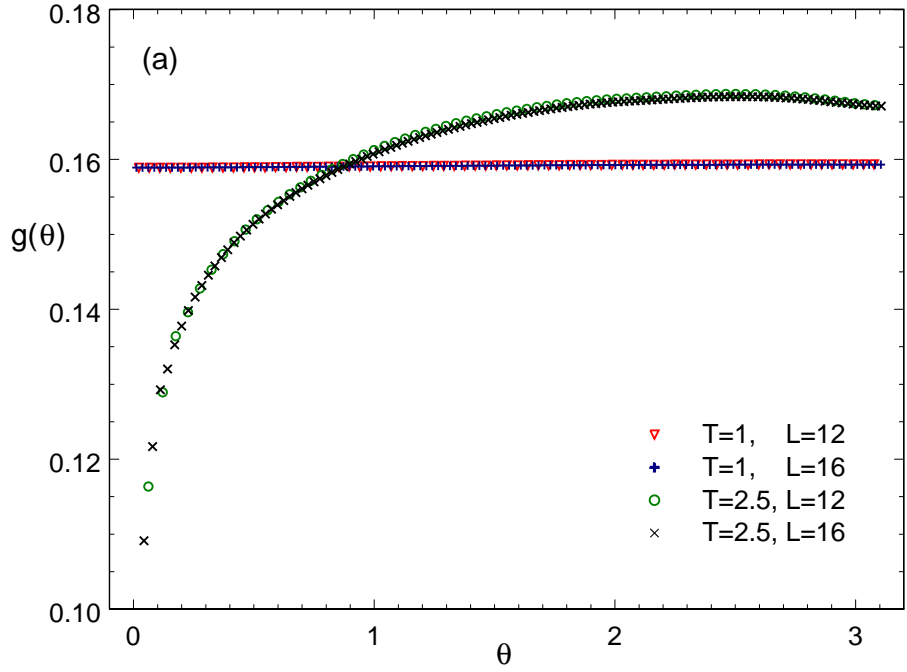}
\includegraphics[width=14cm]{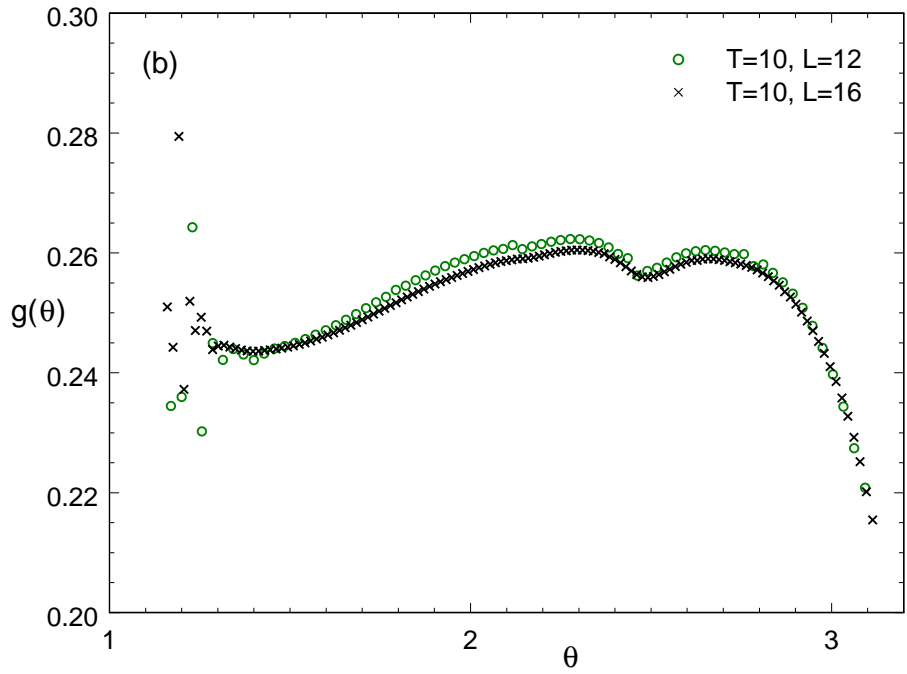}
\caption{(Color online) The density of Yang-Lee zeros for $L=12$ and 16.
(a) $T=1$ and $T=2.5$. (b) $T=10$ (in unit of $J/k_B$).}
\end{figure}

\begin{figure}
\includegraphics[width=9cm]{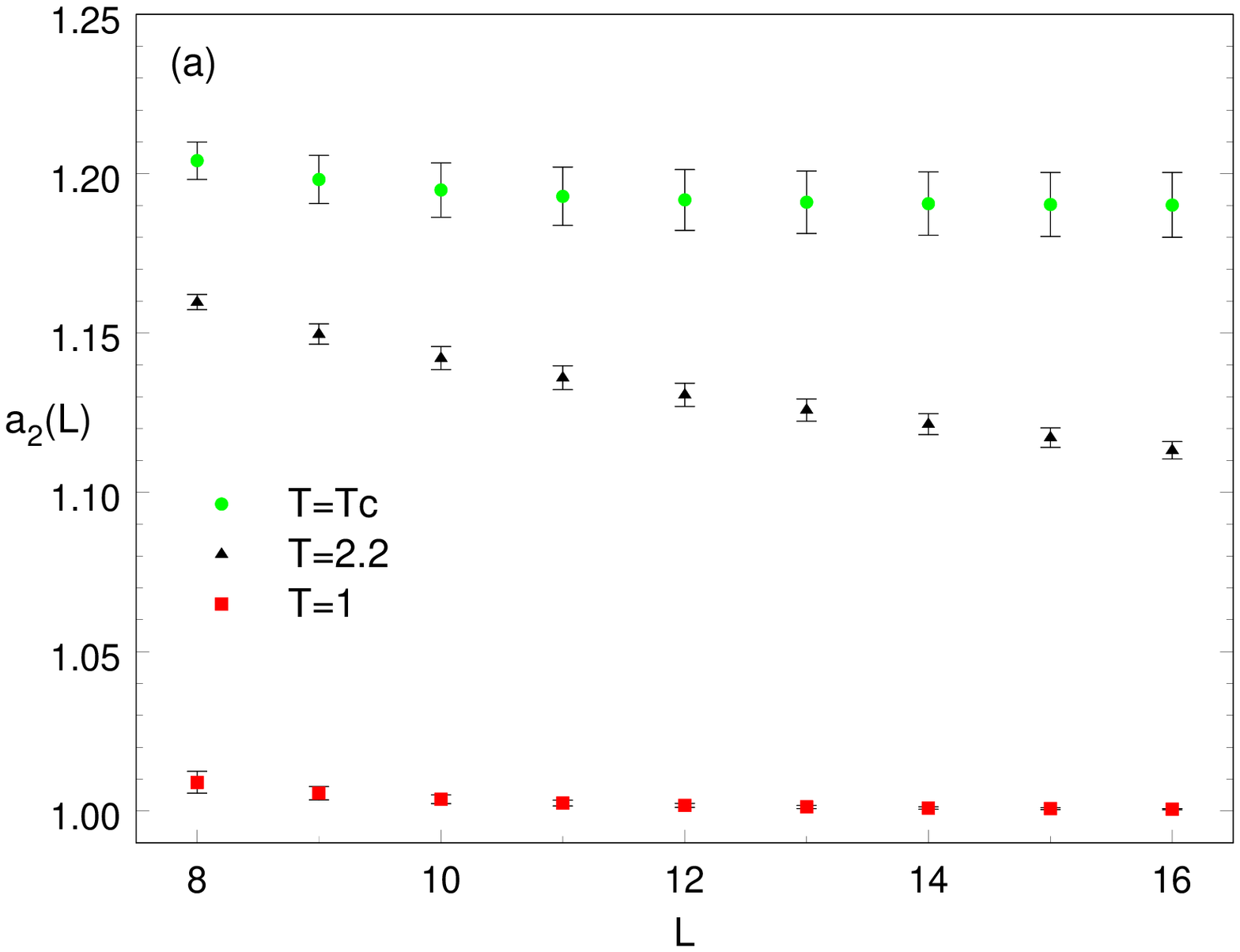}
\includegraphics[width=9cm]{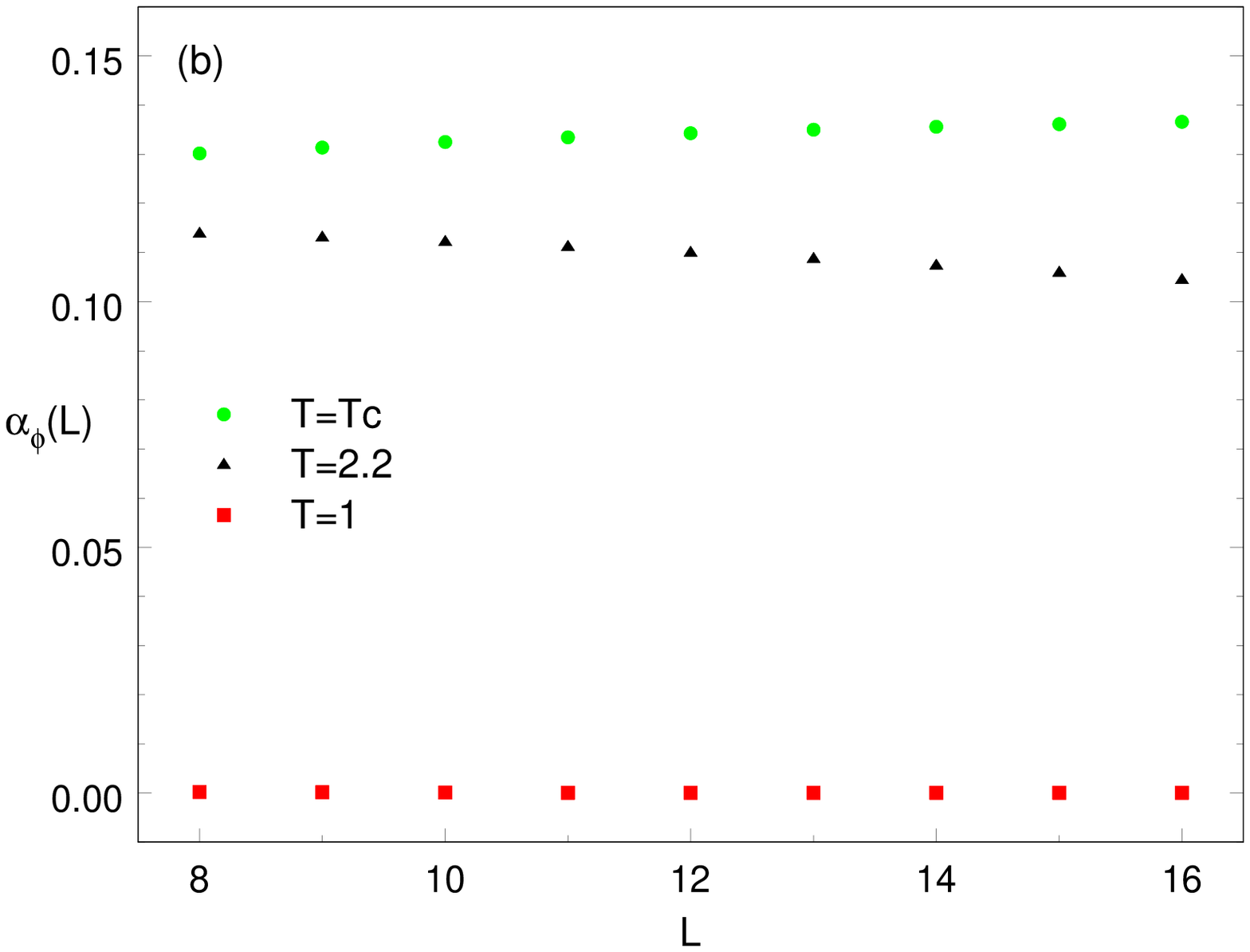}
\includegraphics[width=9cm]{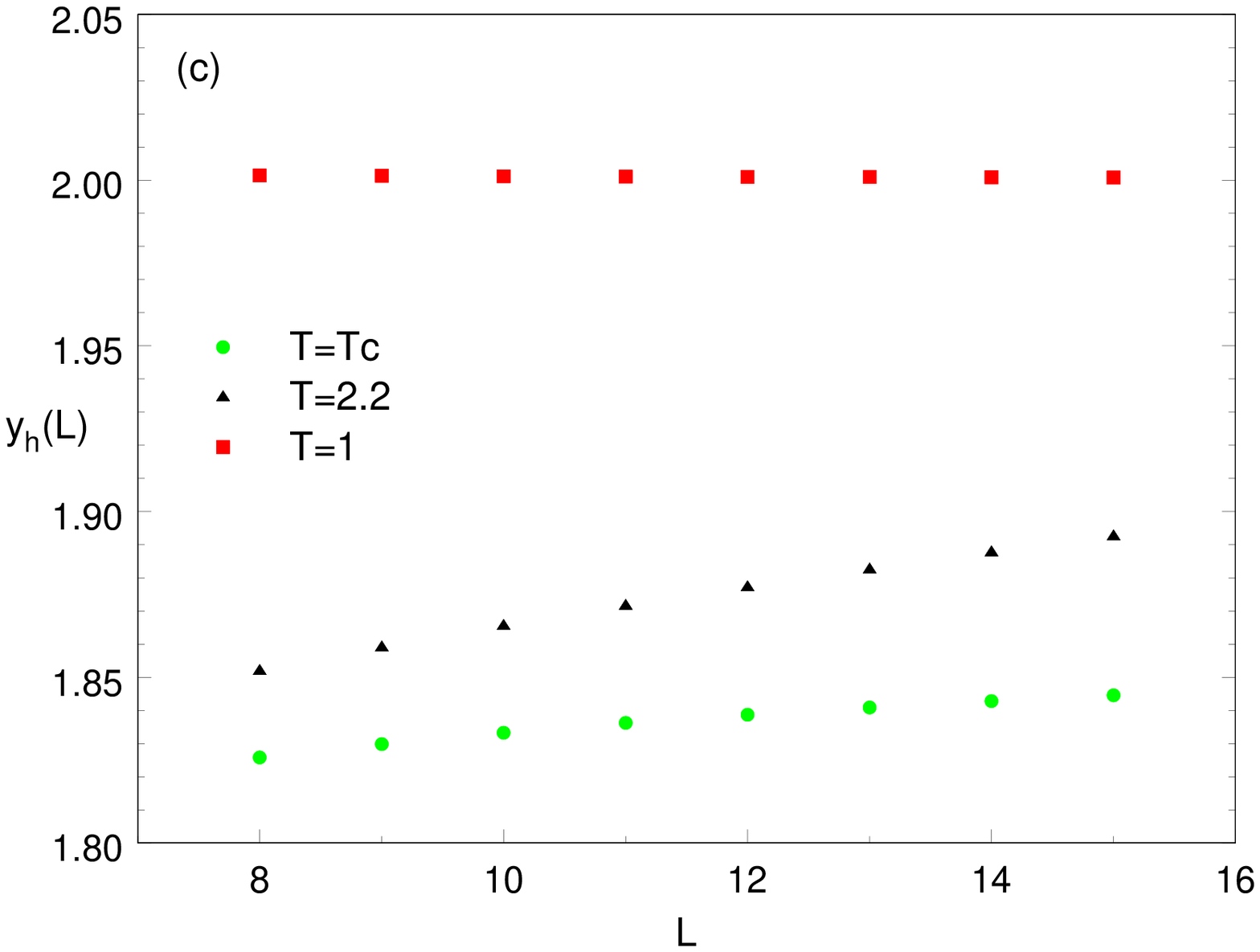}
\caption{(Color online) The values of (a) $a_2(L)$ (JK approach), (b) $\alpha_\phi(L)$ (BMH approach),
and (c) $y_h(L)$ (ADOZ approach), respectively,
for $T=1$, 2.2, and $T_c$ (in unit of $J/k_B$), obtained from Yang-Lee zeros of sizes $L=8$ to 16.}
\end{figure}


\end{document}